\documentclass[prc,twocolumn,showpacs]{revtex4}
\usepackage{graphicx}
\begin{document}

\title{Fusion of radioactive $^{132}$Sn with $^{64}$Ni}
\date{\today}

\author{J.~F.~Liang, D.~Shapira, J.~R.~Beene, C.~J.~Gross,
  R.~L.~Varner, A.~Galindo-Uribarri, J.~Gomez~del~Campo,
  P.~A.~Hausladen, P.~E.~Mueller, D.~W.~Stracener}
\affiliation{Physics Division, Oak Ridge National Laboratory, Oak Ridge,
Tennessee 37831}
\author{H.~Amro, J.~J.~Kolata}
\affiliation{Department of Physics, University of Notre Dame, Notre Dame,
IN 46556}
\author{J.~D.~Bierman}
\affiliation{Physics Department AD-51, Gonzaga University, Spokane, Washington
99258-0051}
\author{A.~L.~Caraley}
\affiliation{Department of Physics, State University of New York at Oswego,
Oswego, NY 13126}
\author{K.~L.~Jones}
\affiliation{Department of Physics and Astronomy, Rutgers University,
Piscataway, NJ 08854}
\author{Y.~Larochelle}
\affiliation{Department of Physics and Astronomy, University of Tennessee,
Knoxville, Tennessee 37966}
\author{W.~Loveland, D.~Peterson}
\affiliation{Department of Chemistry, Oregon State University, Corvallis,
Oregon 97331}

\begin{abstract}
Evaporation residue and fission cross sections of radioactive
$^{132}$Sn on $^{64}$Ni were measured near the Coulomb barrier.
A large sub-barrier fusion
enhancement was observed. Coupled-channel calculations including
inelastic excitation of the projectile and target, and neutron transfer
are in good agreement with the measured fusion excitation function.
When the change in nuclear size and shift in barrier height are
accounted for, there is no extra fusion enhancement in
$^{132}$Sn+$^{64}$Ni with respect to stable Sn+$^{64}$Ni.
A systematic comparison of evaporation residue cross sections for
the fusion of even $^{112-124}$Sn and $^{132}$Sn with $^{64}$Ni is presented.  
\end{abstract}


\maketitle

\section{Introduction}
Fusion of heavy ions has been a topic of interests for several
decades\cite{re94}. One motivation is to understand the reaction
mechanisms so that the production yield of heavy elements can be
better estimated by model calculations. The formation of a compound nucleus is
a complex process. The projectile and target have to be captured
inside the Coulomb barrier and subsequently evolve into a compact
shape. In heavy systems, the dinuclear system can separate during
shape equilibration prior to passing the saddle point. This
quasifission process is considered the primary cause of fusion
hindrance\cite{ba85,to85,hi05}.

At energies near and below the Coulomb barrier, the
structure of the participants plays an important role in influencing
the fusion cross section\cite{be88,da98,ba98}.
Sub-barrier fusion enhancement due to nuclear
deformation and inelastic excitation has been
observed\cite{le95,bi96,mo94,st95,so98}.
Coupled-channel calculations have successfully reproduced
experimental data by including nuclear deformation and inelastic
excitation. Nucleon transfer is another important channel to be
considered\cite{st95r,ti98}.

Recently available radioactive ion beams offer the
opportunity to study fusion under the influence of strong nucleon
transfer reactions. Several theoretical works have predicted large
enhancement of sub-barrier fusion involving neutron-rich radioactive
nuclei\cite{ta92,hu91,da92,de00,za03}.
In addition, the compound nucleus produced in such reactions
is predicted to have a higher survival probability and longer lifetimes.
This is
encouraging for superheavy element research. If high-intensity, 
neutron-rich radioactive beams become available in the future, new
neutron-rich heavy nuclei may be synthesized with enhanced
yields. The longer lifetime of new isotopes of heavy elements would
enable the study of their atomic and chemical properties\cite{ho01}. However,
the current intensity of the radioactive beams is several orders of
magnitude lower than that of stable beams. It is thus not practical to use
such beams for heavy element synthesis experiments, but they do provide
excellent opportunities for studying reaction mechanisms of fusion
involving neutron-rich radioactive nuclei.

Fusion enhancement, with respect to a one-dimensional barrier
penetration model prediction, has been observed in experiments performed with
neutron-rich radioactive ion beams at sub-barrier
energies\cite{ko98,wa01,zy97,li03,li05}. For
instance, the effect of large neutron excess on fusion enhancement
can be seen in $^{29,31}$Al+$^{197}$Au\cite{wa01}.
However, when comparing reactions involving stable isotopes of the projectile
or target, the fusion excitation functions are very similar if the
change in nuclear sizes is accounted for.

This paper reports results of fusion excitation functions measured with
radioactive $^{132}$Sn on $^{64}$Ni. The doubly magic (Z=50, N=82)
$^{132}$Sn has eight neutrons more than the heaviest stable $^{124}$Sn.
Its N/Z ratio (1.64) is larger than that of stable doubly magic nuclei
$^{48}$Ca (1.4) and $^{208}$Pb (1.54) which are commonly used for
heavy element production\cite{ho00}. Evaporation residue (ER) and fission
cross sections were measured. The sum of ER
and fission cross sections are taken as the fusion cross section.

The experimental apparatus is described in Sect. II and data
reduction procedures in Sect. III. The results and
comparison with model calculations are presented in Sect. IV. In
Sect. V a comparison of ER and fusion cross sections with those resulting
from stable Sn
isotopes on $^{64}$Ni is discussed. A summary is given in Sect. VI.
 
\section{Experimental Methods}
The experiment was carried out at the Holifield Radioactive Ion Beam
Facility. A~42 MeV proton beam produced by the Oak Ridge Isochronous
Cyclotron was used to bombard a uranium carbide target. The fission
fragments were ionized by an electron beam plasma ion source. The
largest yield of mass A=132 fragments was $^{132}$Te. Therefore,
it was necessary to suppress $^{132}$Te. This was accomplished by
introducing sulfur into the ion source then selecting the mass 164 XS$^{+}$
molecular ions from the extracted beam. The $^{132}$Te to $^{132}$Sn ratio
in the ion beam was found to be suppressed by a large factor
($\sim 7\times10^{4}$) compared
to that observed with the mass 132 atomic beam. The mass 164 SnS$^{+}$
beam was converted into a Sn$^{-}$ beam by passing it through a Cs vapor
cell where the molecular ion underwent breakup and charge
exchange\cite{st03}. The negatively
charged Sn was subsequently injected into the 25~MV electrostatic tandem
accelerator
to accelerate the beam to high energies. The measurement was
performed at energies between 453 and 620 MeV.
The average beam intensity was 50,000 particles per
second (pps) with a maximum of 72,000 pps. The ER cross sections
measured between 453 and 560 MeV have been reported previously\cite{li03}.

\begin{figure}[ht]
\centering
\includegraphics[width=3.25in]{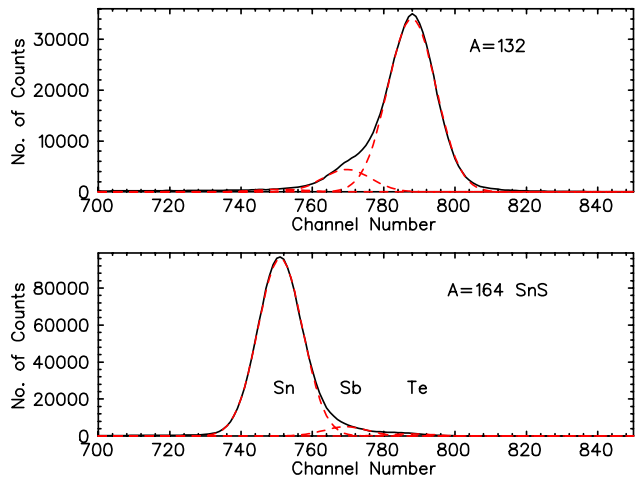}
\caption{(Color online) Composition of a 560 MeV mass A=132 beam
  measured by the ionization chamber. Top panel: The mass A=132
  beam without purification where Te and Sb are the major components
  of the beam. Bottom panel: Sulfur was introduced into the target
  ion source and SnS was selected by the mass separator.
  The dashed curves are results of fitting the spectrum with
  three Gaussian distributions. The isobar contaminants $^{132}$Sb and
  $^{132}$Te were suppressed considerably. \label{fg:puresn}}
\end{figure}

The purity of the Sn beam was measured by an ionization chamber
mounted at zero degrees. Figure~\ref{fg:puresn} displays the energy loss
spectra of a 560~MeV A=132 beam with and without the sulfur
purification. The dashed curves are the results of fitting the
spectrum with Gaussian distributions to estimate the composition of
the beam. In the upper panel, the beam is primarily $^{132}$Te without
sulfur in the ion source. When sulfur was introduced in the ion source,
the beam was 96\% $^{132}$Sn, as shown in the lower
panel. The small amount of Sb and Te had a negligible impact on the
measurement because their atomic number is higher. Fusion of the target
with these isobaric contaminants at sub-barrier energies
should have been suppressed due to the higher Coulomb barriers.

\begin{figure}[ht]
\centering
\includegraphics[width=3.25in]{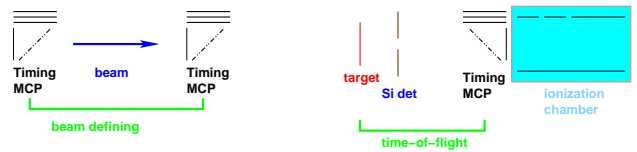}
\caption{(Color online) Apparatus for measuring fission and
  evaporation residues cross sections induced by low intensity beams in
  inverse kinematics.\label{fg:setup}}
\end{figure}

The apparatus for the fusion measurement is shown in
Fig.~\ref{fg:setup}. A thick $^{64}$Ni target (1.0 mg/cm$^{2}$) was
used to compensate for the low beam intensity. Since the compound nucleus
decays by particle evaporation and fission, the evaporation residue
(ER) and fission cross sections were measured. The ERs were
detected by the ionization chamber at zero degrees
and the fission fragments were detected by an annular double-sided
silicon strip detector.

The ERs were identified by the time-of-flight measured with the
microchannel plate timing detector located
in front of the ionization chamber and by
energy loss in the ionization chamber. The two microchannel plate timing
detectors located before the target were used to monitor the beam
intensity and to provide the timing reference for the
time-of-flight measurement.
The microchannel plate timing detector in front of the ionization
chamber was position sensitive and was used to monitor the beam
position. It was located 200~mm from the target and had a 25~mm diameter
Mylar foil. The ionization chamber was filled with
CF$_{4}$ gas so that it could function
at rates up to 50,000 pps. Higher beam intensities occurred in some of
the fission measurements, requiring the ionization chamber to be turned off.
The data acquisition was triggered by either
the beam signal rate down scaled by a factor of 1000, the coincidence of
the delayed beam signal and ER signal, or the silicon detector signal.
A 350~MeV Au beam that resembled ERs was measured by the
ionization chamber to calibrate the energy
loss spectrum. The ER cross section was obtained by taking the ratio of the
ER yield to the target thickness and the integrated beam particles
in the ionization chamber.
A detailed description of the ER measurement technique
used in this experiment can be found in Ref.~\cite{sh05}.

The annular double-sided silicon strip detector (Micron Semiconductor Design
S2) was located 42~mm from the
target. It had
48 concentric strips on one side and 16 pie-shaped sectors on the
other side. The inner diameter was 35~mm and the outer diameter was 70~mm.
The thickness of the detector was 300~$\mu$m.
The detection angles spanned 15.6$^{\circ}$ to 39.6$^{\circ}$.
The fission fragments were identified by requiring a coincidence of events
in the Si detector and by the folding angle distributions of the detected
particles.

\section{Data reduction procedures}

\subsection{Evaporation residues}
Since this was an inverse kinematics reaction, the ERs recoiled in
the forward direction in a narrow cone. The apparatus was designed to
have high efficiency for detecting ERs.
The efficiency of the apparatus was estimated by Monte Carlo
simulations. The angular distribution of the ERs was generated by
statistical model calculations using the code {\tt
PACE2}\cite{ga80}. The input parameters for the statistical model
calculations will be discussed later in this paper. The calculated
efficiency for the lowest bombarding energy is 93$\pm$1\%. It increases as
the reaction energy increases and reaches 98$\pm$1\% at the highest
energy.

A relatively thick target was used in this experiment. The beam lost
approximately 40~MeV after passing through the target
(13~MeV in the center of mass). For this reason, the measured cross
section is an average
of the contributions from the beam interacting throughout the
thickness of the target. The variation of ER cross sections is not
very large at energies above the Coulomb barrier because the
shape of the excitation function is almost flat. Therefore, the
measured cross section is close to that would be measured at an energy
corresponding to the middle of the target. However, at energies below
the barrier
the ER cross section falls off exponentially. The cross section near the
entrance of the target has more weight than that near the exit.
Smooth curves fitting the excitation function in this rapidly varying
region were used to
determine the reaction energy associated with the measured cross section.

An iterative method was used to determine the effective
reaction energy for the thick
target measurement . First, the measured cross sections and the beam
energies calculated at the middle of the target were fitted by a
tensioned spline\cite{cl81} where the smoothness of the curve could be
adjusted. The
resulting curve was then used to calculate the thick target cross section 
for each measurement, according to
\[ \sigma_{i} = \int \frac{\sigma(E)}{dE/dx} dE/\rho \]
where $\sigma(E)$ is the curve generated by the spline fit, $dE/dx$
is the stopping power of $^{132}$Sn in $^{64}$Ni, and $\rho$ is the target
thickness. The integration limits were the energies of the beam at the
exit of the target and at the entrance of the target. The energy, $E_{i}$,
corresponding to the cross section, $\sigma_{i}$, was obtained by
interpolation using the fitted curve. This set of energies was used as the
input for the next iteration of the fit. The result converged very quickly.
After five iterations, the energies differed from the previous
iteration energies by less than 0.2 MeV. The validity of this
method was checked by generating data from a known
function such as the Wong formula~\cite{wo73} and folding in the
effects of target thickness.

\begin{figure}[ht]
\centering
\includegraphics[width=3.25in]{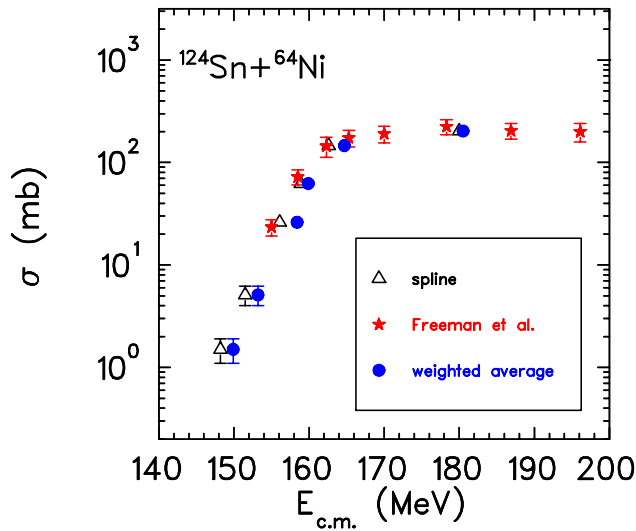}
\caption{(Color online) Comparison of ER cross sections for
  $^{64}$Ni+$^{124}$Sn measured in this work and by Freeman {\it et
  al.}\protect{\cite{fr83}} (filled stars). The filled circles are for
  energies determined by the method described in Ref.~\protect{\cite{sh05}}
  and the open triangles are for energies determined by the method described
  in this paper.\label{fg:sn124ni}}
\end{figure}

Comparing to the cross-section-weighted-average method described in
Ref.~\cite{sh05}, the differences in energies determined by these two
methods are not noticeable at high energies
because the excitation function is fairly flat. However, at energies below
the barrier, the energy determined by the cross-section-weighted-average
method
is larger than that determined by the method described above and
disagrees with the measurement in Ref.~\cite{fr83}, as can be seen in
Fig.~\ref{fg:sn124ni}. Furthermore, it
is found that using data generated from a known function the effective
energy obtained by the cross-section-weighted-average method is shifted
to too high an energy in the exponential falloff region.

The uncertainty of the energy
determination was estimated by comparison with the method using the
cross section weighted average. The average uncertainty of the effective
reaction energy is 2.3 ~MeV in the region where the excitation
function is almost flat and increases to 3.9~MeV in the exponential fall off
region. The uncertainty is larger, 5.8~MeV, for the
lowest energy data point because an extrapolation is required for
calculating the thick target cross section and the extrapolation
region is influenced by the location of the next higher energy point.

To verify our measurement technique, the ER cross sections for
$^{124}$Sn+$^{64}$Ni in inverse kinematics were measured and compared
to those published by Freeman {\it et al.} measured with a thin
target\cite{fr83}. It is noted that some of our measurements were
performed at energies different from those of Ref.~\cite{fr83}. 
The comparison is shown in Fig.~\ref{fg:sn124ni}.
Our data (open triangles) are in good agreement with those measured
by Freeman {\em et al.}~\cite{fr83} (filled stars). The solid circles
are for energy determined by the cross-section-weighted-average method
described in Ref.~\cite{sh05}.

\subsection{Fission}
\begin{figure}[ht]
\centering
\includegraphics[bb= 18 198 576 735, width=3.25in]{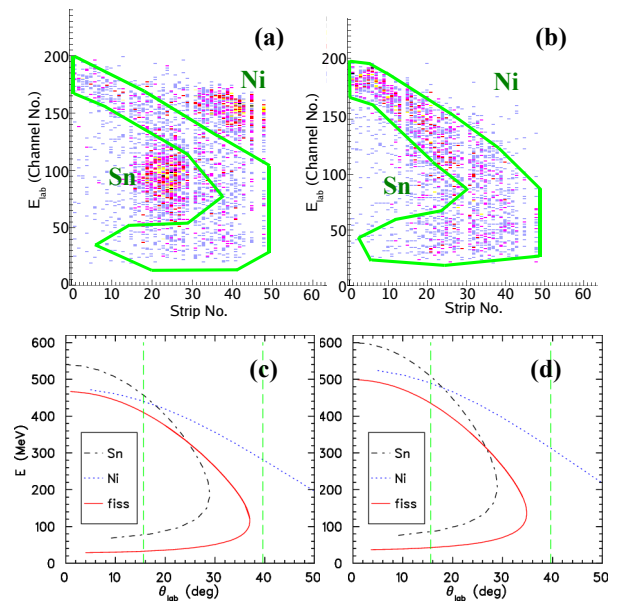}
\caption{(Color online) (a) and (b) Two dimensional histograms of energy and
  strip number for coincident events
  from 560 and 620 MeV $^{132}$Sn+$^{64}$Ni, respectively, measured
  by the annular
  double-sided silicon strip detector. The gated area shows events
  from fission and other reactions. (c) and (d) Kinematics of energy as a
  function of scattering angle for 560 and 620 MeV $^{132}$Sn+$^{64}$Ni,
  respectively, elastic scattering and fission fragments.The dash-dotted
  and dotted curves are for the elastically scattered Sn and Ni,
  respectively whereas the solid curve is for the fission
  fragments. The angular range of the Si strip detector is between the
  two vertical dashed lines.\label{fg:evsth}}
\end{figure}

Fission fragments were identified by requiring a coincidence of two
particles detected by the pie-shaped sectors of the Si strip detector
on either side of the beam. Figure~\ref{fg:evsth}(a) and (b) present
two-dimensional histograms of particle energy and strip number of the Si
detector for coincident events taken from 560~MeV and 620~MeV
$^{132}$Sn+$^{64}$Ni, respectively. They were
compared to the kinematics calculation displayed in
Fig.~\ref{fg:evsth}(c) and (d) where the fission fragments,
elastically scattered
Sn and Ni are shown by the solid, dash-dotted and dotted curves,
respectively. The angular range of the Si strip detector is between
the two vertical dashed lines. The elastically scattered
Ni and Sn appear in the upper right hand corner and center of the histogram,
respectively. The fission events are located in the gated area.

\begin{figure}[ht]
\centering
\includegraphics[width=3.25in]{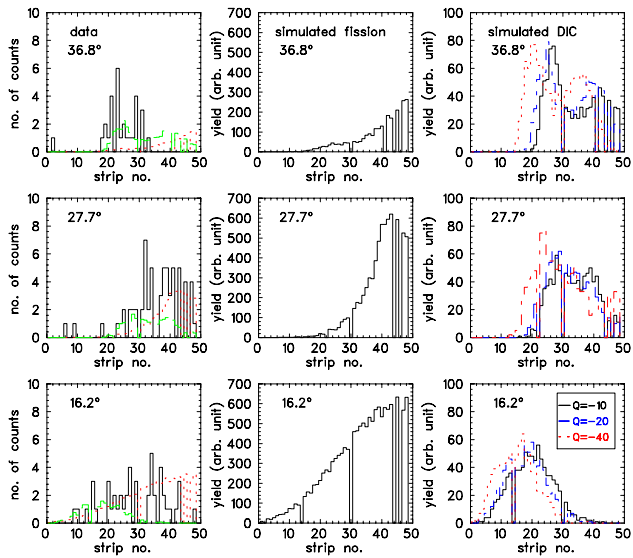}
\caption{(Color online) Left panels: Folding angle distributions for
  550 MeV $^{124}$Te+$^{64}$Ni for one of the fragments detected
  at 16.2$^{\circ}$ (strip 2), 27.7$^{\circ}$ (strip 22), and
  36.8$^{\circ}$ (strip 41) by the annular double-sided silicon strip detector.
  The elastic scattering events are excluded. The
  dotted and dashed histograms are the results of fitting the data
  with simulated fission and deep inelastic collisions with Q=--20
  MeV, respectively (see text). Middle
  panels: Results of Monte Carlo simulations for fission
  events. Right panels: Results of Monte Carlo simulations for deep
  inelastic scattering events. The solid curves are for reaction Q
  value of --10 MeV, the dashed curves are for Q=--20 MeV, and the
  dotted curves are for Q=--40 MeV.\label{fg:tefold}}
\end{figure}

The folding angle distributions of the fragments were used to distinguish
fission from other reactions, such as deep inelastic reactions.
Since there are two solutions for the
kinematics of the inverse reaction, as shown in
Fig.~\ref{fg:evsth}(c) and (d), the fragment angular correlation is not as
simple as that in normal kinematics. Monte Carlo simulations were
performed to provide guidance. It was assumed that only fusion-fission
results from a full momentum transfer. The width of the mass
distribution was taken from the $^{58}$Ni+$^{124}$Sn
measurement\cite{wo87}. The width of the mass distribution was varied
to estimate the uncertainty of the simulation. The transition state
model\cite{va73} was used to predict the fission fragment angular
distribution. In Fig.~\ref{fg:tefold} the simulated fission fragment folding
angle distributions for 550 MeV $^{124}$Te+$^{64}$Ni are compared
with a stable beam test measurement.
The folding angle distributions for one of the
fragments detected in strip 2 (16.2$^{\circ}$), strip 22 (27.7$^{\circ}$), and
strip 41 (36.8$^{\circ}$) are shown. The gap in the spectra at strip
14, 30, 44, 46, and 47 are malfunctioning strips in the detector.

The Monte Carlo simulated folding angle distributions for fission are
shown in the middle panels of Fig.~\ref{fg:tefold} and compared to
those of measurements shown in the left panels. For one of the fragments
detected at forward angles, strip 2 for example, the predicted
angular distribution of the other fragment is
similar to that of the measurement. Most of these events are considered as
resulting from fission. For one of the fragments detected near the middle
part of the detector, strip 22 for instance, there are
differences between measurement and simulation in the shapes
of the angular distributions of the other fragment. It
is predicted that the other fission fragment is distributed around strip 40.
The measured distribution spreads to more forward angles. For one of
the fragments detected at the backward angles, the yield of the other
fragment is predicted to be small and they are equally distributed
between the middle part of the detector and the outer edge of the detector.
But the measured events appear in the middle part of
the detector. There are no events in the region where fission events
are expected. These differences are attributed to the contribution
from other reaction mechanisms, most likely deep inelastic collisions.

An attempt was made to simulate these deep inelastic collision
events. It was assumed that the mass of these products were
projectile- and target-like and the angular distribution
at forward angles followed a 1/sin($\theta$) dependence. The right
panels of Fig.~\ref{fg:tefold} show the results of simulations
performed for reaction Q values of --10 (solid), --20 (dashed), and
--40 MeV (dotted). It can be
seen that the overlap of fission and deep inelastic collisions becomes
larger at more backward angles. At
strip 41 (36.8$^{\circ}$), deep inelastic collisions account for all the
events.

The relative contribution of fission and deep inelastic collisions
were obtained by fitting the simulated folding angle distributions
to the measured distributions for all the detector strips using
the CERN library program {\tt MINUIT}\cite{ja03}. In the fits, the
normalization coefficients for the simulated distributions were the
only two variable parameters. The results of the fits are shown in the left
panels of Fig.~\ref{fg:tefold} by the dotted and dashed histograms
for fission and deep inelastic collisions with Q=--20 MeV,
respectively. The number of fission events in the measured
distributions were taken as the summed events in each strip
multiplied by the relative contribution of fission.

\begin{figure}[ht]
\centering
\includegraphics[width=3.25in]{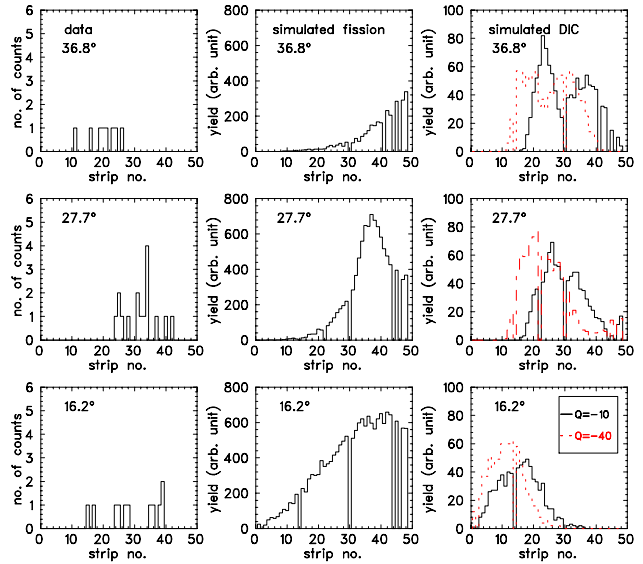}
\caption{(Color online) Left panels: Folding angle distributions for
  560 MeV $^{132}$Sn+$^{64}$Ni for one of the fragments detected
  at 16.2$^{\circ}$ (strip 2), 27.7$^{\circ}$ (strip 22), and
  36.8$^{\circ}$ (strip 41) by the annular double-sided silicon strip detector.
  The elastic scattering events are excluded. Middle
  panels: Results of Monte Carlo simulations for fission
  events. Right panels: Results of Monte Carlo simulations for deep
  inelastic scattering events. The solid curves are for reaction Q
  value of --10 MeV, and the
  dotted curves are for Q=--40 MeV.\label{fg:snfold}}
\end{figure}

The folding angle distributions for $^{132}$Sn+$^{64}$Ni are shown in
Fig.~\ref{fg:snfold}. Due to the low statistics, it was not practical
to extract the fission events by fitting the folding angle
distributions. As an alternative, the fission events were extracted by
setting gates on the folding angle distributions using the simulated
distributions as references. This gating method was also tested with the
$^{124}$Te+$^{64}$Ni measurement. The fission cross sections obtained
by the fitting method and the gating method agreed within 10\%.

The Monte Carlo simulation was also employed to calculate the coincidence
efficiency of the
detector. The efficiency increased from 5.7$\pm$0.9\% at 530 MeV
to 7.6$\pm$0.8\% at 620 MeV bombarding energy.

In the present work, the dynamic range of the amplifiers was not sufficiently
large resulting in the distortion of the high energy signals.
In the future, new amplifiers that are more
suitable for measuring the energy of fission fragments will be used so that
the mass ratio of reaction products can be obtained to help distinguish
fission events from other reaction channels.

The formation of a compound nucleus depends on whether the interacting
nuclei are captured inside the fusion barrier and  whether the dinuclear
system can subsequently evolve into a compact shape. Quasifission
occurs when the dinuclear system fails to cross the saddle point to
reach shape equilibrium. 
Since the beam intensity was several orders of magnitude lower than that of
stable beams and the reaction was in
inverse kinematics, making separation of fusion-fission and quasifission
very difficult, there was no attempt to distinguish quasifission
from fusion-fission in this work. Furthermore, the experimental
results are compared to barrier penetration models which describe the
capture process, making it unnecessary to separate these two
processes.

\section{Comparison with model calculations}
\subsection{Statistical model}
\begin{figure}[ht]
\centering
\includegraphics[width=3.25in]{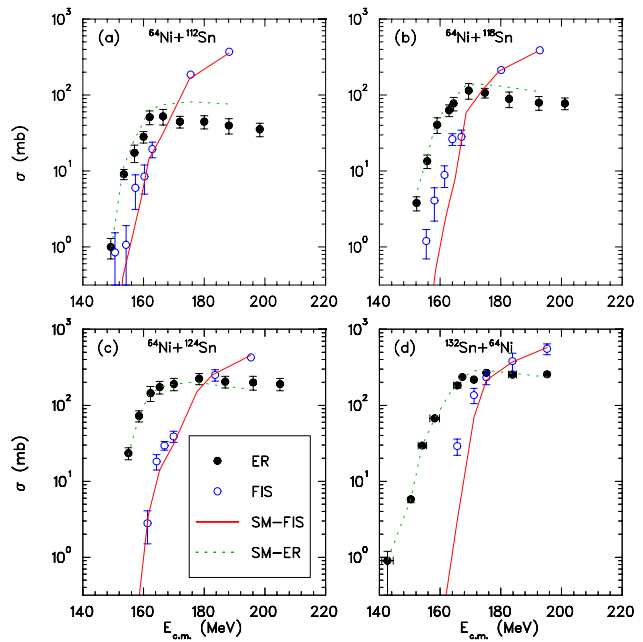}
\caption{(Color online) Comparison of measured ER (filled circles)
  and fission (open
  circles) cross sections with statistical model calculations. The
  solid and the dotted curves are the calculated fission and ER cross
  sections, respectively, using the measured fusion cross sections as
  input. (a) $^{64}$Ni+$^{112}$Sn, (b) $^{64}$Ni+$^{118}$Sn,
  (c) $^{64}$Ni+$^{124}$Sn (Freeman {\it et al.}\protect{\cite{fr83}}),
  and (d) $^{132}$Sn+$^{64}$Ni (this work) \label{fg:snni-sm}}
\end{figure}

The compound nucleus formed in $^{132}$Sn+$^{64}$Ni decays by particle
evaporation and fission. Statistical models have
successfully described compound nucleus decay for a wide range of
fusion reactions. The measured ER and fission cross sections
are compared with the predictions of
the statistical model code {\tt PACE2}\cite{ga80}.
The input parameters were
obtained by simultaneously fitting the data from stable Sn on
$^{64}$Ni\cite{fr83,le86} and the measured fusion cross
sections\cite{le86} were used for the calculations.
Figure~\ref{fg:snni-sm}(a), (b), and (c)
displays the comparison of calculations and data for
$^{112,118,124}$Sn+$^{64}$Ni, respectively. The calculations
reproduce the measurements well except for the ER cross sections
of $^{112}$Sn+$^{64}$Ni. Table~\ref{tb:snni-sm} lists the input
parameters for the calculations. Without adjusting the parameters,
calculations for $^{132}$Sn+$^{64}$Ni were performed. The results are
shown in Fig.~\ref{fg:snni-sm}(d). Very good agreement between the
calculation and the data can be seen.
\begin{table}
\begin{tabular}{lc}\hline\hline
level density parameter (a)  & A/8 MeV$^{-1}$\\
a$_{f}$/a$_{n}$  & 1.04 \\
diffuseness of spin distribution ($\Delta l$)       & 4 $\hbar$ \\
fission barrier  & Sierk\cite{si86} \\ \hline\hline
\end{tabular}
\caption{Input parameters for statistical model calculations.
 \label{tb:snni-sm}}
\end{table}

It is noted that some of the parameters used in our calculations are
different from those used by Lesko {\em et al.}~\cite{le86}. In their
calculations, the code {\tt CASCADE}\cite{pu77} was used.
The mass of the nuclei
in the decay chain was calculated using the Myers droplet
model\cite{my77}. The diffuseness of the spin distribution was $\Delta l = 15
\hbar$  and the ratio of level density at the saddle point to the
ground state, $a_{f}/a_{n}$, was set to 1.0. In this work, a
compilation of measured masses\cite{wa03}, $\Delta l = 4 \hbar$, and
$a_{f}/a_{n} = 1.04$ were used.

\subsection{Coupled-channel calculation}
\begin{figure}[ht]
\centering
\includegraphics[width=3.25in]{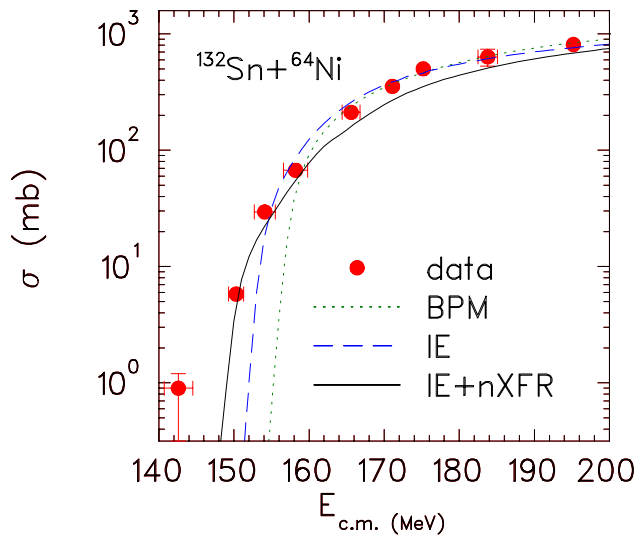}
\caption{(Color online) Comparison of $^{132}$Sn+$^{64}$Ni fusion
  data (filled circles) with a one-dimensional barrier penetration model
  calculation (dotted curve). The coupled-channel calculation
  including inelastic excitation of the projectile and target is shown
  by the dashed curve and the calculation including inelastic excitation
  and neutron transfer is shown by the solid curve.
  \label{fg:snni-cc}}
\end{figure}

In general, sub-barrier fusion enhancement can be described by
coupled-channel calculations. The fusion cross section of
$^{132}$Sn+$^{64}$Ni, the sum of ER and fission cross sections,
is compared with coupled-channel calculations using the code {\tt
CCFULL}\cite{ha99}. The interaction potential (V$_{\circ}$=82.46~MeV,
r$_{\circ}$=1.18~fm, and a=0.691~fm) was taken from the systematics of
Broglia and Winther\cite{br91}. The result of the calculations are
compared with the data in Fig.~\ref{fg:snni-cc}. The dotted curve is
the prediction of a one-dimensional barrier penetration model and
it can be seen that it
substantially underpredicts the sub-barrier cross sections. The
coupled-channel calculation including inelastic excitation of
$^{64}$Ni to the first 2$^{+}$ and 3$^{-}$ states and
$^{132}$Sn to the first 2$^{+}$ state is shown by the dashed
curve. The transition matrix elements, B(E$\lambda$), of $^{64}$Ni
were obtained from Ref.~\cite{ra87,sp89} and the B(E2) of $^{132}$Sn
was obtained from a recent measurement by Varner {\it et
al.}\cite{va05}. This calculation overpredicts the data at energies near the
barrier and underpredicts the data well below the barrier.

The neutron
transfer reactions have positive Q values for transferring two to six
neutrons from $^{132}$Sn to $^{64}$Ni. Since there is no  neutron
transfer data available for this reaction, the transfer coupling form
factor is unknown. Thus, the coupled-channel calculation including transfer
and inelastic excitation was performed with one effective transfer
channel using the Q value for two-neutron transfer.
The coupling constant was adjusted to fit the data. The
calculation with the coupling constant set to 0.48 is shown by the
solid curve. It reproduces the data very well except for the lowest
energy data point which has large uncertainties in energy and in cross
section. A better treatment of the transfer channels based on experimental
transfer data would help improve
understanding of the influence of transfer on fusion.
Experimental neutron transfer data on
$^{132}$Sn+$^{64}$Ni in the future would be very useful.
 
\section{Discussion}
\begin{figure}[ht]
\centering
\includegraphics[width=3.25in]{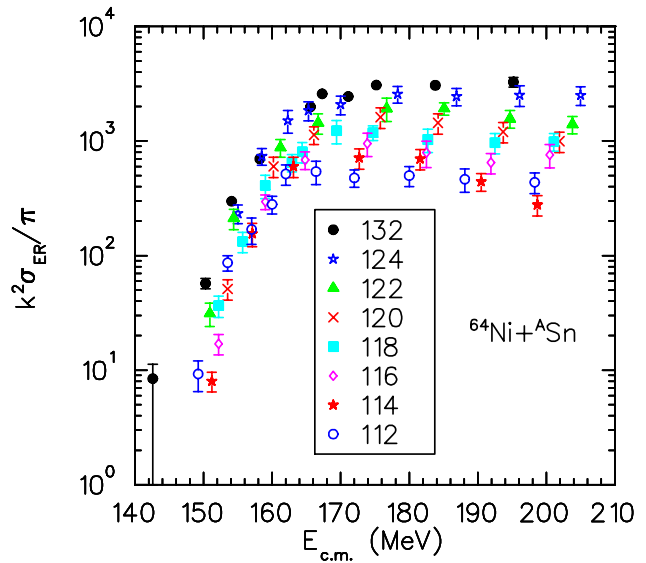}
\caption{(Color online) The reduced ER cross section as a function
  of the center of mass
  energy for $^{64}$Ni on stable even Sn isotopes\protect{\cite{fr83}}
  and radioactive $^{132}$Sn.\label{fg:rervsex}}
\end{figure}

The ER cross section can be described by
\[ \sigma_{ER} =
\pi\lambdabar^{2}\sum_{l=0}^{l_{c}} (2l+1)\sigma_{l}, \]
where $\lambdabar$ is the de Broglie wave length, $l_{c}$ the maximum
angular momentum for ER formation and $\sigma_{l}$ the partial cross
section. The reduced ER cross sections for $^{64}$Ni on stable-even Sn
isotopes\cite{fr83} are compared with that for $^{132}$Sn+$^{64}$Ni in
Fig.~\ref{fg:rervsex}. The reduced ER cross section is defined as the
ER cross section divided by the kinematic factor
$\pi\lambdabar^{2}$. It can be seen that the ER cross sections saturate
at high energies as fission becomes a significant fraction of the fusion
cross section. In addition, the saturation value increases as the
neutron excess in Sn increases. This is consistent with the fact that the
fission barrier height increases for the more neutron-rich
compound nuclei.

\begin{figure}[ht]
\centering
\includegraphics[width=3.25in]{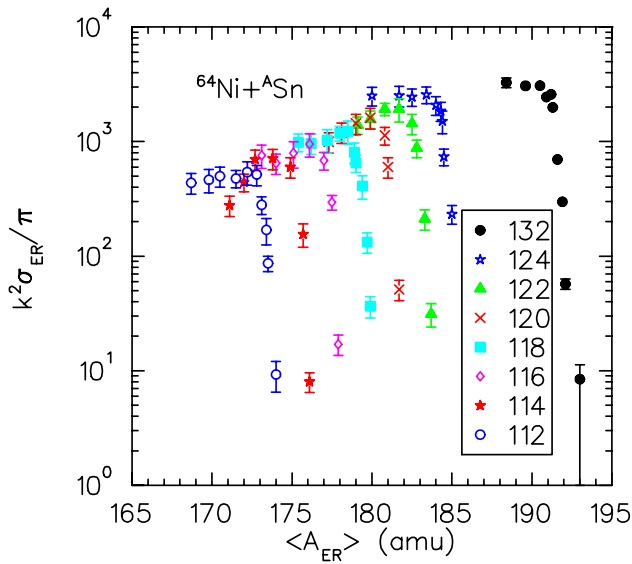}
\caption{(Color online) The reduced ER cross section as a function
  of the calculated
  average mass of ERs predicted by {\tt PACE2}\protect{\cite{ga80}}
  for $^{64}$Ni on stable even Sn isotopes\protect{\cite{le86}}
  and radioactive $^{132}$Sn. \label{fg:rervsa}}
\end{figure}

In Fig.~\ref{fg:rervsa}, the measured reduced ER cross sections for Ni+Sn
as a function of the calculated average mass of the ERs,
predicted by {\tt PACE2},
are presented. In the same reaction, the higher mass ERs are produced
at lower beam energies because of the lower excitation energies of the
compound nucleus. As the neutron excess in the compound nucleus
increases, neutron evaporation becomes the dominant decay channel.
The {\tt PACE2} calculation predicts that a
compound nucleus made with Sn isotopes of mass
number greater than 120 decays essentially 100\% by neutron evaporation
and Pt isotopes are the primary ERs. The mass of the compound
nucleus is different when it is produced with different Sn isotopes.
However, it can be seen that Pt of a particular mass can be produced
with different Sn isotopes if different numbers of neutrons are
evaporated. The reaction with a more neutron-rich Sn
produces the same Pt isotope at a higher rate. With $^{132}$Sn as the
projectile, the ERs are so neutron-rich that they cannot be produced by
stable Sn induced reactions. This suggests that it
may be beneficial to use neutron-rich radioactive ion beams to produce
new isotopes of heavy elements.

\begin{figure}[ht]
\centering
\includegraphics[width=3.25in]{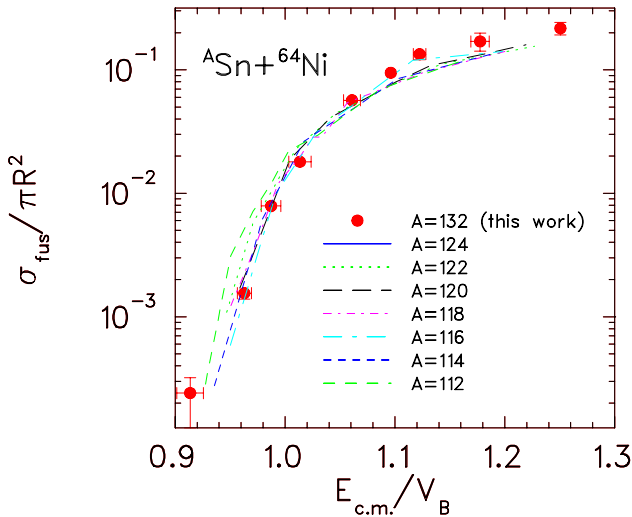}
\caption{(Color online) Comparison of fusion excitation functions
  for $^{64}$Ni on
  stable even Sn isotopes\protect{\cite{le86}} and radioactive
  $^{132}$Sn. The change in nuclear sizes are
  corrected by factoring out the area and the Bass barrier
  height\protect{\cite{ba74}} in
  the cross section and energy, respectively. \label{fg:snniall}}
\end{figure}

The fusion excitation functions of $^{64}$Ni on stable even Sn
isotopes\cite{le86} are compared with that of $^{132}$Sn+$^{64}$Ni in
Fig.~\ref{fg:snniall}. In order to remove the effects of the difference in
nuclear sizes, the cross section is divided by $\pi$R$^{2}$
with R=1.2(A$_{p}^{1/3}$+A$_{t}^{1/3}$)~fm, where A$_{p}$ (A$_{t}$)
is the mass number of the projectile (target). The reaction energy in the
center of mass is divided by the barrier height predicted by the Bass
model\cite{ba74}. It can be seen that the fusion of $^{132}$Sn and
$^{64}$Ni is not
enhanced with respect to the stable-even Sn isotopes when the
difference in nuclear sizes is considered.

The lowest energy data point has large uncertainties. The cross
section seems enhanced comparing to the stable beam measurements in
Fig.~\ref{fg:rervsex} and Fig.~\ref{fg:snniall}. A more pronounced
enhancement appears when the data point is compared to our
coupled-channel calculations (Fig.~\ref{fg:snni-cc}) and to a
time-dependent Hartree-Fock calculation\cite{um06}. To further
explore if fusion is enhanced at this low energy region, we plan
to repeat the measurement with an improved apparatus where the
thickness of the Mylar foil in the microchannel plate timing detector
located in front of the ionization chamber will be reduced.
This will allow a better separation of the energy loss signals from ERs
and scattered beams in the ionization chamber at low bombarding
energies.

The Q values for transferring two to six neutrons from $^{132}$Sn to
$^{64}$Ni are positive. It is necessary to include neutron transfer in
coupled-channel calculations to reproduce experimental results.
As the neutron excess in the Ni isotopes
decreases, the number of neutron transfer channels with positive Q
values increases for $^{132}$Sn+Ni. In $^{132}$Sn+$^{58}$Ni, the Q
values for
transferring one to sixteen neutrons from $^{132}$Sn to $^{58}$Ni are
positive and range from 1.7 to 17.4 MeV. A large sub-barrier fusion
enhancement due to the coupling to neutron transfer is expected to
occur in $^{132}$Sn+$^{58}$Ni. An experiment to measure the fusion
excitation function of $^{132}$Sn on $^{58}$Ni is in preparation.

Although $^{132}$Sn is unstable, its neutron separation energy is 7.3
MeV.
This is not very low compared to stable nuclei. The sub-barrier fusion
enhancement observed in $^{132}$Sn+$^{64}$Ni with respect to stable Sn
nuclei can be accounted for by the change in nuclear sizes. No extra
enhancement was found. However, an increased ER yield at energies
above the barrier was observed as compared to stable Sn. As the shell
closure is crossed, the binding energy for $^{133}$Sn decreases by a
factor of two. The nuclear surface of $^{133}$Sn and even more neutron-rich
Sn may be more diffused. The number of neutron transfer channels
with positive Q values increases by a factor of two or more. Larger
sub-barrier fusion enhancement beyond the nuclear
size effect may be expected.

\section{Summary}
Neutron-rich radioactive $^{132}$Sn beams were incident on a $^{64}$Ni target
to measure fusion cross sections near the Coulomb barrier. With an
average intensity of 5$\times$10$^{4}$ pps beams and a high efficiency
apparatus for ER detection, the uncertainty of the measured ER cross
section is small and comparable to that achieved in stable beam
experiments. The efficiency for fission fragment detection was
low but the detector had a very fine granularity. By requiring a
coincident detection of the fission fragments and performing folding
angle distribution analysis, fission events were identified. The
excitation functions of ER and fission can be described by statistical
model calculations using parameters that simultaneously fit the
stable even Sn isotopes on $^{64}$Ni fusion data.
A large sub-barrier fusion enhancement with respect to
a one-dimensional barrier penetration model prediction was
observed. The enhancement is attributed to the coupling of the
projectile and target inelastic excitation and neutron transfer. The
reduced ER cross sections at energies above the barrier are larger for the
$^{132}$Sn induced reaction than those induced by stable Sn
nuclei, as expected from the higher fission barrier of the more neutron-rich
compound
nucleus. For a specific mass of ER, reactions with a more neutron-rich
Sn have higher cross sections. When the fusion excitation functions
are compared on a reduced scale, where the effects of nuclear size and
barrier height are
factored out, no extra fusion enhancement is observed in
$^{132}$Sn+$^{64}$Ni with respect to stable Sn induced fusion. The
fusion cross section measured at the lowest energy seems to be
enhanced. Experiments to investigate this with an improved apparatus
is planned.

\section{Acknowledgment}
We would like to thank D. J. Hinde for helpful and stimulating discussions.
We wish to thank the HRIBF staff for providing excellent radioactive beams and
technical support. Research at the Oak Ridge National Laboratory is
supported by the U.S. Department of Energy under contract DE-AC05-00OR22725
with UT-Battelle, LLC. W.L. and D.P. are supported by the the U.S. Department
of Energy under grant no. DE-FG06-97ER41026.

\end{document}